\documentclass[epj]{svjour}

\usepackage{graphicx}
\usepackage{fancyhdr}
\usepackage{color}
\def\makeheadbox{{
 }}
\def\mytitle{My title} 
\def\myauthors{My name}  
\def\mytype{My type of session}
\def\mysession{My session}

\def\mytitle{Domain Wall Networks} 
\def\myauthors{Norisuke Sakai}    
\def\mytype{Contributed Talk}    
\def\mysession{Theoretical Models}

\begin{document}
\title{Effective Lagrangian of Domain Wall Networks}
\author{Norisuke~Sakai\inst{1}
\thanks{\emph{Email: nsakai@th.phys.titech.ac.jp} 
}%
 \and
 Minoru~Eto\inst{1}
\thanks{\emph{Present address:} INFN, Sezione di Pisa,
Largo Pontecorvo, 3, Ed. C, 56127 Pisa, Italy
}%
 \and
 Toshiaki~Fujimori\inst{1}
 \and
 Takayuki~Nagashima\inst{1}
 \and
 Muneto~Nitta\inst{2}
 \and
 Keisuke~Ohashi\inst{1}
\thanks{\emph{Present address:} Department of Applied 
Mathematics and Theoretical Physics, 
University of Cambridge, CB3 0WA, UK
}%
}                     
%
%
\institute{Department of Physics, Tokyo Institute of
Technology, Tokyo 152-8551, Japan
\and 
Department of Physics, 
Keio University, Hiyoshi, Yokohama, Kanagawa 223-8521, Japan
}
%
\date{}
\abstract{
Domain wall networks are studied in ${\cal N}=2$ 
supersymmetric $U(N_{\rm C})$
gauge theory with $N_{\rm F}(>N_{\rm C})$ flavors. 
We find a systematic method to construct domain wall 
networks in terms of moduli matrices. 
Normalizable moduli parameters of the network are found to 
be sizes and phases of the loop. 
We obtain moduli space metric which specifies the effective 
Lagrangian on the domain wall networks. 
It is used to study dynamics of domain wall networks with 
the moduli approximation. 
} 
\maketitle
\section{Introduction}
\label{intro}
D-branes play an essential role in understanding 
nonperturbative dynamics in string theories. 
As an extended object preserving half of the 
supersymmetry (SUSY), 
they have many similarities with domain walls in field 
theories. 
More than two domain walls 
intersect or meet with angles in general, 
and networks or webs of these objects emerge 
when many domain walls meet at various junctions. 
Junctions and networks of domain walls have many 
similarities with those of D-branes. 
Both of them can preserve a quarter of SUSY. 
Therefore they are called $1/4$ BPS states. 
A few exact solutions of domain wall junctions have been 
obtained some time ago \cite{Oda:1999az}. 

The purpose of the present paper is to give a compact 
summary of our study of domain wall networks in $U(N_{\rm C})$ 
gauge theories with $N_{\rm F} (> N_{\rm C})$ Higgs scalars 
in the fundamental representations. 
1/2 BPS parallel walls can exist for 
hypermultiplets with real masses 
in supersymmetric theories 
with eight supercharges in spacetime 
dimensions $d \leq 5$  
\cite{Tong:2002hi,Isozumi:2004jc}. 
On the other hand, 
to obtain non-parallel walls, we need complex masses 
for hypermultiplets 
which is possible in $d \leq 4$.
We find that genuine moduli corresponding to normalizable 
modes are sizes of wall loops and their associated phases. 
We obtain 
the 
metric of these moduli in the effective Lagrangian 
of domain wall networks. 
The effective Lagrangian allows us to discuss dynamics 
of domain wall networks in the moduli approximation. 

More detailed analysis of some topics on domain wall 
networks may be found in our recent papers 
\cite{Eto:2005cp}, 
and a general survey of the moduli matrix approach is 
given in our review \cite{Eto:2006pg}. 
As a review for other solitons in the same model, 
see \cite{Tong:2005un}.

\section{$1/4$ BPS Equations}
\label{sec:BPSequations} 
\subsection{SUSY $U(N_{\rm C})$ Gauge Theory 
with $N_{\rm F}$ Flavors
}
\label{sec:gaugetheory} 




We consider 3+1 dimensional  ${\cal N} = 2$ supersymmetric 
$U(N_{\rm C})$ gauge theory with $N_{\rm F}(>N_{\rm C})$ 
massive hypermultiplets in the fundamental representation. 
The physical bosonic fields contained in this model are a gauge 
field $W_\mu\ (\mu=0,1,2,3)$ and real adjoint scalars 
$\Sigma_\alpha\ (\alpha=1,2)$ 
in the vector multiplet, and $N_{\rm F}$ complex 
doublets of scalars $H^{iAr}$ 
$(r=1,2,\cdots,N_{\rm C},\ A=1,2,\cdots,N_{\rm F},\ i=1,2)$ 
in the hypermultiplets.
We express $N_{\rm C}\times N_{\rm F}$ matrix of the 
hypermultiplets by $H^i$. 
With the metric $\eta_{\mu\nu}=(+1,-1,-1,-1)$, we obtain 
the bosonic Lagrangian with the gauge coupling $g$ 
\begin{eqnarray}
{\cal L} 
&=&
{\rm Tr}\biggl[
-\frac{1}{2g^2}F_{\mu\nu}F^{\mu\nu}
+ \frac{1}{g^2}\sum_{\alpha=1}^2
{\cal D}_\mu\Sigma_\alpha{\cal D}^\mu\Sigma_\alpha 
\nonumber \\
&&
+ {\cal D}_\mu H^i\left({\cal D}^\mu H^i\right)^\dagger
\biggr] - V,
\label{lag}\\
V &=& {\rm Tr}\biggl[
\sum_{\alpha=1}^2\left(H^i M_\alpha - \Sigma_\alpha H^i\right)
\left(H^i M_\alpha - \Sigma_\alpha H^i\right)^\dagger
\label{pot}
\\
&-& 
\left.\frac{1}{g^2}\left[\Sigma_1,\Sigma_2\right]^2
+ 
\frac{g^2}{4}
\left(H^{1}H^{1\dagger}-H^{2}H^{2\dagger}-c\mathbf{1}_{N_{\rm C}}\right)^2
\right],
\nonumber 
\end{eqnarray}
with diagonal mass matrices 
$M_1={\rm diag}\left(m_1,m_2,\cdots,m_{N_{\rm F}}\right)$ and 
$M_2 = {\rm diag}\left(n_1,n_2,\cdots,n_{N_{\rm F}}\right)$,  
and $c>0$ the Fayet-Iliopoulos (FI) parameter. 
The covariant derivatives and the field strength are defined by
 ${\cal D}_\mu \Sigma_\alpha = \partial_\mu \Sigma_\alpha 
+ i[W_\mu,\Sigma_\alpha]$, 
 ${\cal D}_\mu H^i = \partial_\mu H^i + iW_\mu H^i$ and 
 $F_{\mu\nu}=
 \partial_\mu W_\nu -\partial_\nu W_\mu +i[W_\mu, W_\nu]$,  
respectively. 

\subsection{Vacua and BPS Equations
}
\label{sec:vacua} 

Supersymmetric vacuum is characterized 
by a set of $N_{\rm C}$ different flavor indices 
$\langle A_1A_2\cdots A_{N_{\rm C}}\rangle $
out of $N_{\rm F}$ flavors, which correspond to 
the color component of nonvanishing hypermultiplet 
\cite{Arai:2003tc}
\begin{eqnarray}
&& H^{1rA}=\sqrt{c}\,\delta ^{A_r}{}_A,\quad H^{2rA}=0, \\
&&\Sigma 
\equiv 
\Sigma_1+i\Sigma_2 
\\
&=& {\rm diag}\left(m_{A_1}+in_{A_1},
\cdots,
m_{A_{N_{\rm C}}}+in_{A_{N_{\rm C}}}\right). 
\nonumber
\end{eqnarray}
This is the Higgs phase, where only the domain walls and 
vortices can exist as elementary solitons. 
Monopoles and instantons can exist when they accompany 
vortices to form composite solitons 
\cite{Tong:2005un,Eto:2006pg}.

By assuming dependence on $x^1, x^2$ and 
requiring $1/4$ of supercharges to be conserved, 
we find the $1/4$ BPS equations for the web of walls 
\begin{eqnarray}
\!\! \!\left[{\cal D}_1+\Sigma_1,{\cal D}_2+\Sigma_2\right]=0,&& 
{\cal D}_\alpha H = HM_\alpha - \Sigma_\alpha H,
\label{bps_eq1}
 \\
\sum_\alpha{\cal D}_\alpha\Sigma_\alpha &=&
\frac{g^2}{2}\left(c{\bf 1}_{N_{\rm C}} - HH^\dagger\right),
\label{bps_eq2}
\end{eqnarray}
with $\alpha=1,2$. 
Solutions of the $1/4$ BPS equations saturate the lower bound 
of the energy $E$ of the field configurations, which is given by 
the sum of topological charges ($\alpha$ in $Z_\alpha$ is not summed) 
\begin{eqnarray}
&&E \ge Z_1+Z_2+Y, 
\quad 
{
Z}_\alpha \equiv \int d^2 x c \partial_\alpha 
{\rm Tr} \Sigma_\alpha, 
\nonumber \\
&&{
Y} \equiv \int d^2 x \frac{2}{g^2}
\partial_\alpha{\rm Tr}\left(\epsilon^{\alpha\beta}
\Sigma_2{\cal D}_\beta\Sigma_1\right)
\label{eq:bps_bound}
\end{eqnarray}

\subsection{BPS Solutions and Moduli Space 
}
\label{sec:solutions}

The first equation in Eqs.(\ref{bps_eq1}) is the 
integrability condition for the 
second equation in Eqs.(\ref{bps_eq1}), 
whose solutions are obtained in terms of 
$N_{\rm C}\times N_{\rm C}$ 
non-singular matrix $S(x^\alpha)$ as
\begin{eqnarray}
&&H = S^{-1}H_0e^{M_1x^1 + M_2x^2},\;
W_1 - i\Sigma_1 = -iS^{-1}\partial_1 S,
\nonumber \\
&&W_2 - i\Sigma_2 = -iS^{-1}\partial_2 S.
\label{bps_sol}
\end{eqnarray}
Here $H_0$ is an $N_{\rm C}\times N_{\rm F}$ constant complex 
matrix of rank $N_{\rm C}$. 
We call $H_0$ the {\it moduli matrix} because it contains 
moduli parameters of solutions as we see below.
Defining a gauge invariant matrix 
$ \Omega\equiv SS^\dagger$, 
Eq.(\ref{bps_eq2}) can be written as
\begin{eqnarray}
{1 \over cg^2}\left[
\partial_\alpha\left(\partial_\alpha\Omega\Omega^{-1}\right)
\right]
= {\bf 1}_{N_{\rm C}} - \Omega_0\Omega^{-1},
\label{master}
\end{eqnarray}
with 
$
 \Omega_0 \equiv c^{-1}H_0e^{2(M_1x^1+M_2x^2)}H_0^\dagger
$. 
We call Eq.~(\ref{master}) the {\it master equation}, 
and assume the existence and the uniqueness of the 
solution once $\Omega_0$ is given. 

Since the same physical  configurations are realized 
by $(H_0,S)$ and $(H_0',S')$ related by the 
{\it $V$-symmetry} with $V \in GL(N_{\rm C},{\bf C})$ 
\begin{equation}
 H_0 \to H_0' = VH_0,\quad S \to S'=VS,
\label{eq:world-volume-sym}
\end{equation}
we obtain the {\it total} moduli space of the web of domain walls 
as the complex Grassmann manifold 
\begin{eqnarray}
\label{tot_mod}
\! \! \! \! \! \! &&G_{N_{\rm F},N_{\rm C}} \simeq
\{H_0\ |\ H_0\sim VH_0,\ V\in GL(N_{\rm C},{\bf C})\} \nonumber 
\\
\! \! \! \! \! \! &&
\simeq SU(N_{\rm F})/
[SU(N_{\rm F}-N_{\rm C}) \times SU(N_{\rm C}) \times U(1)].
\end{eqnarray}
This space is made by gluing all the topological sectors together 
and is not endowed with a metric \cite{Isozumi:2004jc}.

In strong gauge coupling limit $g^2 \to \infty$, 
Eq.~(\ref{master}) can be algebraically solved.
For instance, in the case of Abelian gauge theory 
($N_{\rm C}=1$) the strong coupling limit gives 
configurations of scalar fields up to gauge symmetry as 
\begin{eqnarray}
H^A = \sqrt{c}\,\frac{H_0^Ae^{m_Ax^1+n_Ax^2} }
{\sqrt{\sum_{B=1}^{N_{\rm F}}
|H_0^B|^2e^{2(m_Bx^1+n_B x^2)}}} .
\label{exact}
\end{eqnarray}

\section{Webs of Walls}
\label{sec:webs_walls} 

\begin{figure}[h]
\begin{center}
\includegraphics[width=50mm]{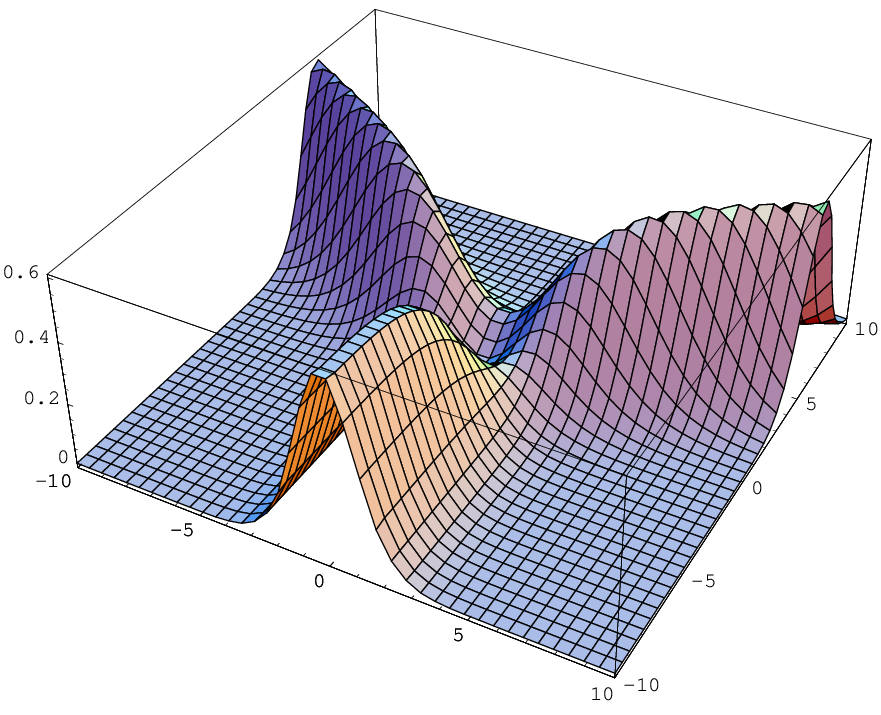}
\caption{Binding energy at the junction 
point:~The energy density is numerically evaluated for 
the moduli matrix 
$ H_0 e^{\vec{m} \cdot \vec{x}} 
= \left( e^{x^2},\, e^{\sqrt{3}x^1/2-x^2/2},
\,e^{ - \sqrt{3}x^1/2-x^2/2}\right)$, 
gauge coupling $g=1$ and FI parameter $c=1$. 
}
\label{fig:junction}
\end{center}
\end{figure}

The moduli matrix of $U(1)$ gauge theory is given by 
\begin{eqnarray}
H_0 = 
\sqrt c (e^{a_1 + ib_1}, \cdots, e^{a_{N_{\rm F}} + ib_{N_{\rm F}}}). 
\end{eqnarray}
In the case of $N_{\rm F} = 2$, two vacua exist and 
a wall connecting them 
is located where two vacuum weights are equal 
\begin{equation}
 (m_1-m_2) x^1 + (n_1-n_2) x^2 + a_1 - a_2 = 0.
\label{wall}
\end{equation}
In the case of $N_{\rm F} = 3$, 
there exist 
3 discrete vacua labeled by $\left<A\right>$ $(A=1,2,3)$, 
and three walls can meet 
to form a junction as illustrated in Fig.\ref{fig:junction}. 
We can recognize that the topological charge $Y$ associated to 
the junction gives a negative contribution to the energy 
density, which can be interpreted as binding energy of domain 
walls. 
This is a feature in the $U(1)$ gauge theory. 
The $1/4$ BPS wall junction is characterized by 
a triangle with 3 vertices at $m_A+in_A$ in $\Sigma$ plane. 
We call such polygons in the $\Sigma$ plane as 
grid diagrams.

\begin{figure}[ht]
\begin{center}
\begin{tabular}{ccc}
\includegraphics[height=2.4cm]{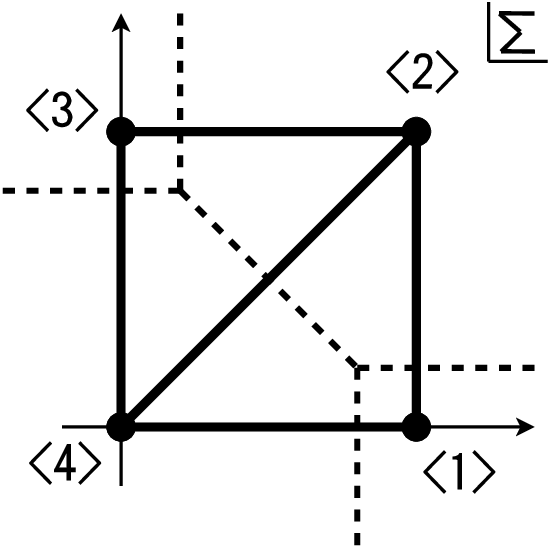} &\quad &
\includegraphics[height=2.6cm]{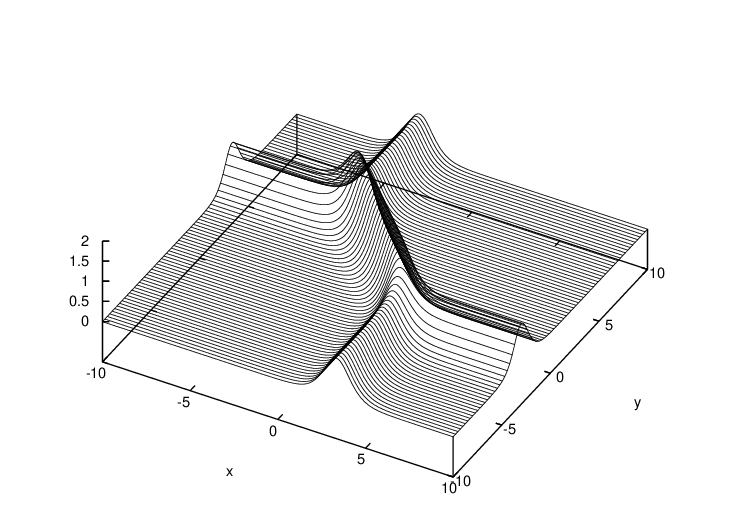} \\
{ (a) grid diagram} &&
{ (b) energy density ($g^2\to\infty$)}
\end{tabular}
\caption{Wall web with 
4 external legs of walls. 
Grid diagram:(a), and 
energy density:(b). 
($[m_A,n_A]=\{[1,0],[1,1],[0,1],[0,0]\}$)}
\label{cp3}
\end{center}
\end{figure}
\begin{figure}[ht]
\begin{center}
\begin{tabular}{ccc}
\includegraphics[height=2.5cm]{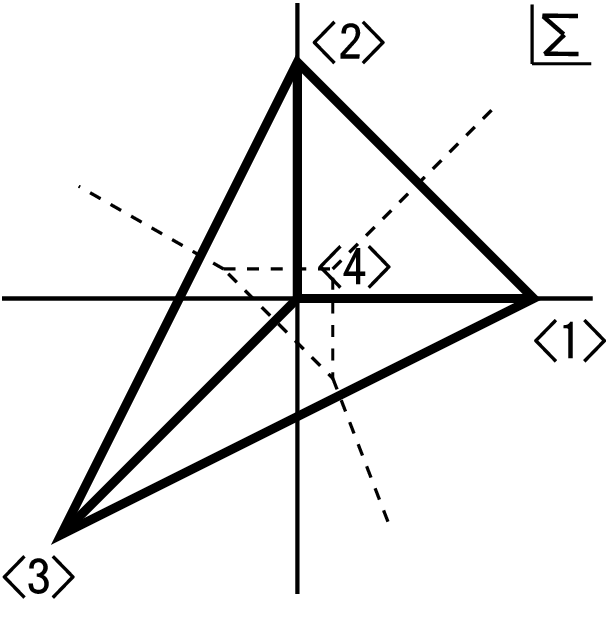} & &
\includegraphics[height=2.5cm]{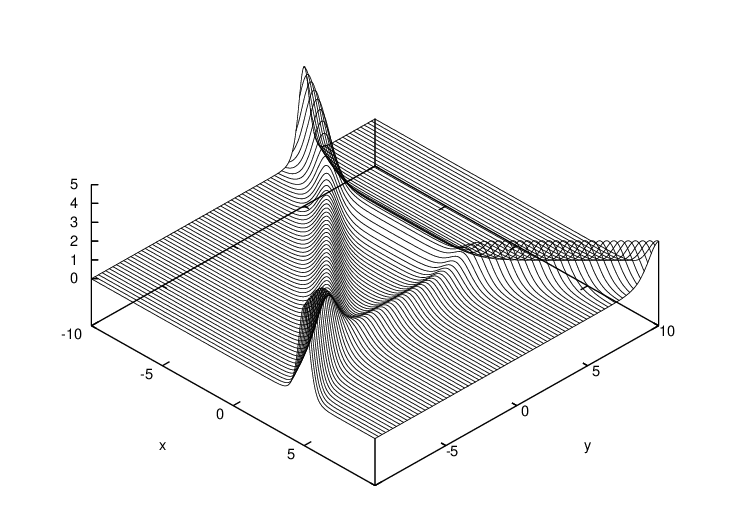}\\
{ (a) grid diagram} &&
{ (b) energy density ($g^2\to\infty$)} 
\end{tabular}
\caption{Web with 
1 loop in the $N_{\rm F}=4$ model. 
Grid diagram:(a), and 
energy density:(b). 
($[m_A, n_A]=[1,0],\ [0,1],\ [-1,-1],\ [0,0]$) 
}
\label{cp3_loop}
\end{center}
\end{figure}

The models with $N_{\rm F} \geq 4$ 
admit more ample webs of walls. 
Let us take the case of 
$N_{\rm F} = 4$ model.
In a choice of mass parameters, we obtain tree diagram 
as shown in Fig.\ref{cp3}. 
In another choice of mass parameters, we obtain a loop 
diagram as shown in Fig.\ref{cp3_loop}. 

In the case of non-Abelian gauge theories, we can obtain 
positive contributions to the topological charge $Y$. 
We find in that case that the junction contains the Hitchin 
system which arises as a dimensionally reduced 
Yang-Mills 
instanton.

\section{Effective Lagrangian of Wall Loops }
\label{sec:eff_lag} 

To obtain effective Lagrangian on the webs of walls, 
we need to fix boundary conditions at asymptotic infinity. 
With that restriction, we can only fluctuate moduli parameters 
associated with the loop of the web. 
Therefore the 
sizes of the loops and their associated phases are the only 
normalizable moduli which can be promoted to fields on 
the web of walls. 
We find a general formula for the effective Lagrangian 
\cite{Eto:2006uw}
\begin{eqnarray}
\mathcal{L}^{eff} 
&=&
K_{ij^*}(\phi,\phi^*)\partial^\mu \phi^i 
\partial_\mu \phi^{j*}, 
\nonumber \\ 
K(\phi,\phi^*) 
&=& 
K_w (\phi,\phi^*) + K_g (\phi,\phi^*) \\
K_w(\phi,\phi^*) 
&\equiv &
\int d^2x\, 
c\,{\rm log det}\Omega, 
 \label{eq:kah_w}
\\ 
K_g(\phi,\phi^*) 
&\equiv & 
\int d^2x\,
\frac{1}{2g^2}{\rm Tr}(\Omega^{-1}\partial_\alpha
\Omega)^2. 
 \label{eq:kah_g}
\end{eqnarray}

Let us take as an example, $N_{\rm F}=4$ case in $U(1)$ gauge 
theory 
with the mass assignment $[m_4, n_4]=[0, 0]$
where we choose the moduli matrix with the complex 
moduli $\phi=e^{r+i\theta}$ 
\begin{equation}
H_0=\sqrt{c}(1,1,1,\phi) .
\end{equation}

\subsection{Metric at Small Loops 
}
\label{sec:small_loop}

We can explicitly evaluate the metric at small loops 
in the strong coupling limit $g^2 \rightarrow \infty$ 
\begin{eqnarray}
K_w &\equiv& c\,\int d^2x 
\left[\,\log\Omega_0-\log\,\tilde{\Omega}_0\,\right] 
\nonumber \\
&=&c\,\int d^2x \,\log\left(
                  1+\frac{|\phi|^2}{\tilde{\Omega}_0}\right), \\
\Omega_0 &=& e^{2\vec m_1 \cdot \vec x}
+e^{2\vec m_2 \cdot \vec x}+e^{2\vec m_3 \cdot \vec x}+|\phi|^2, \\
\tilde \Omega_0 &\equiv&
   e^{2\vec m_1\cdot \vec x} + e^{2\vec m_2\cdot \vec x} 
 + e^{2\vec m_3\cdot \vec x}, 
\end{eqnarray}
where the mass of the hypermultiplet of the $A$-th flavor is 
denoted by a two-vector $\vec m_A=(m_A, n_A)$, 
 $\vec m_A \cdot \vec x \equiv m_A x^1 + n_A x^2$, and 
$\Delta_{[123]}$ is the area of the triangle in field 
space with the masses of $1,2,3$ as vertices. 

Let us define 
$\alpha_i\equiv {(\vec m_j\times \vec m_k)}/{\Delta_{[123]}}$. 
We find that we can expand the integrand in powers of 
$|\phi|$ in the region of 
 $|\phi|^2 
\le \exp \left( -\sum \alpha_i\log\alpha_i \right)$. 
Therefore we know that there exists well-defined smooth 
function even for 
$|\phi|^2 \ge \exp \left( -\sum \alpha_i\log\alpha_i \right)$. 
Moreover, this can be written as a sum of hypergeometric functions 
in cases where ${\alpha_i}$ 
are rational numbers. 
\begin{figure}[h]
\begin{center}
\includegraphics[width=40mm]{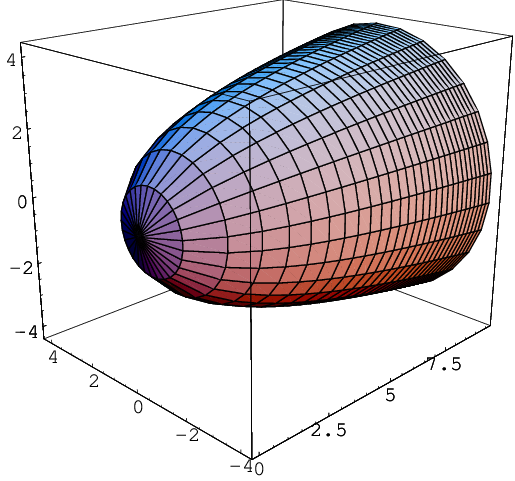}
\end{center}
\caption{The moduli space of single triangle loop 
around $\phi=0$ where the loop shrinks. 
$U(1)$ isometry is 
the phase modulus. 
The other direction 
is the size modulus of the loop. 
}
\label{fig:metric}
\end{figure}
We obtain the K\"ahler potential for $|\phi|^2 
\le \exp \left( -\sum \alpha_i\log\alpha_i \right)$ 
\begin{eqnarray*}
K_w
=
\frac{c}{4\Delta_{[123]}}
\sum_{n=1}^\infty \frac{(-1)^{n+1}}{n}\frac{\Gamma(\alpha_1n)
\Gamma(\alpha_2n)\Gamma(\alpha_3n)}{\Gamma(n)}|\phi|^{2n}.
\end{eqnarray*}
We find that the scalar curvature is finite (nonsingular) 
even at $\phi=0$ (vanishing loop)
\begin{eqnarray}
R 
= \frac{16\Delta_{[123]}}{c} \frac{\Gamma(2 \alpha_1)
\Gamma(2 \alpha_2)\Gamma(2 \alpha_3)}{\left(\Gamma(\alpha_1)
\Gamma(\alpha_2)\Gamma(\alpha_3)\right)^2} 
+ \mathcal O(|\phi|^2).
\end{eqnarray}

\subsection{Metric at Large Loops 
}
\label{sec:large_loop} 

For large sizes of the loop, $\Omega$ in the integrand of 
the K\"ahler potential (\ref{eq:kah_w}) and (\ref{eq:kah_g}) 
is dominated by the vacuum weight in each region, 
apart from finite regions near walls and junctions. 
Therefore evaluation in terms of vacuum weights is 
a good approximation at large values of $|\phi|$. 
This is the so-called tropical limit, and the approximation 
is valid for arbitrary values of gauge coupling $g$. 
By using the tropical approximation, we find 
$K_w$ is given by $c$ times the volume of the tetrahedron 
\begin{eqnarray}
&&K_w^{trop}
= 
\frac{c}{24\Delta_{[123]} }
\frac{1}{\alpha_1 \alpha_2 \alpha_3}(\log|\phi|^2)^3,  
\label{eq:kahlerw}
\\
&&K_g^{trop}=-\frac{2}{g^2}|\vec m_1|^2A_1
-\frac{2}{g^2}|\vec m_2|^2A_2-\frac{2}{g^2}|\vec m_3|^2A_3 
\nonumber \\
&=&
-\frac{(\log |\phi|^2)^2}{4g^2\Delta_{[123]}}
\left(\frac{|\vec m_{12}|^2}{\alpha_3}
+\frac{|\vec m_{23}|^2}{\alpha_1}
+\frac{|\vec m_{31}|^2}{\alpha_2}\right).
\label{eq:kahlerg}
\end{eqnarray}

Combining the above K\"ahler potentials (\ref{eq:kahlerw}) 
and (\ref{eq:kahlerg}), 
we obtain the total K\"ahler metric as 
\begin{eqnarray}
&&ds^2=
\frac{c}{\Delta_{[123]}}\biggl[ \frac{r}{\alpha_1 \alpha_2 \alpha_3}
\\
&-&\frac{1}{g^2 c}\left(\frac{|\vec m_{12}|^2}{\alpha_3}
+\frac{|\vec m_{23}|^2}{\alpha_1}
+\frac{|\vec m_{31}|^2}{\alpha_2}\right)\biggr](m^2dr^2+d\theta^2).
\nonumber
\end{eqnarray}
We can understand the effective action as the 
kinetic energies of walls and junctions due to the 
moduli motion.

\section{Dynamics of Loops}
\label{sec:dynamics}

Let us finally comment on the dynamics of domain wall loops 
by using the effective Lagrangian with the moduli 
approximation. 
The fan-like metric in Fig.\ref{fig:metric} implies that 
the domain wall loops tend to expand without limit. 
Non-Abelian gauge theory 
allows different types of loops 
which can be deformed to each other 
by changing a modulus. 
In this case, 
the moduli geometry looks like a sandglass 
made by gluing the tips of the two cigar-(cone-)like metrics 
of a single triangle loop. 

Then the sizes of all loops tend 
to grow for a late time in general models 
with complex Higgs masses, 
while the sizes are stabilized at some values 
once triplet masses are introduced for the Higgs fields.

Dynamics of a double loop in Abelian gauge theory and 
a non-Abelian loop is discussed in the last paper 
of \cite{Eto:2005cp}.

\section{Conclusion}
\label{sec:conclusion}
\begin{enumerate}
\item
Webs of domain walls are constructed as $1/4$ BPS states 
in ${\cal N}=2$ SUSY $U(N_{\rm C})$ non-Abelian gauge theories 
in $4$ dimensions with 
$N_{\rm F} 
(> N_{\rm C})
$ 
hypermultiplets in the fundamental representation. 
\item
Total moduli space of the webs of walls is given 
by a complex Grassmann manifold $G_{N_{\rm F},N_{\rm C}}$
described by the moduli matrix $H_0$. 
\item
Exact solutions of webs of walls 
are obtained for $g^2\rightarrow \infty$. 
\item
Abelian junction has a topological charge contributing 
negatively to the energy (binding energy). 
Non-Abelian junction has a topological charge contributing 
positively to energy. 
The positive charge is the result of the Hitchin system 
sitting at the junction. 
\item
A general formula for the effective Lagrangian is obtained. 
\item
Normalizable moduli of web of walls are sizes of the loop 
and their associated phases. 
\item
Metric of a single triangle loop of walls is 
explicitly worked out and the dynamics of loops 
are worked out. 
\end{enumerate}


%
%

\end{document}